\documentclass[a4paper]{article}

\usepackage[english]{babel}
\usepackage[utf8x]{inputenc}
\usepackage[T1]{fontenc}

\usepackage[a4paper,top=3cm,bottom=2cm,left=3cm,right=3cm,marginparwidth=1.75cm]{geometry}

\usepackage{amsmath}
\usepackage{graphicx}
\usepackage[colorinlistoftodos]{todonotes}
\usepackage[colorlinks=true, allcolors=blue]{hyperref}

\title{Quantum Network Recovery from Multinode Failure using Network Encoding with GHZ-States on Higher-Order Butterfly Networks}
\author{Mrittunjoy Guha Majumdar and Shayan Srinivasa Garani}

\begin{document}
\maketitle

\begin{abstract}
We propose a protocol to transmit three quantum states crossly in a butterfly
network with prior entanglement, in the form of GHZ states, between three senders. The proposed
protocol requires only one qubit transmission or two classical bits transmission in each channel
of the network. We generalize this protocol to higher number of qubits with multiqubit GHZ
states towards quantum network operability using network coding with multiqubit GHZ states on
higher-order butterfly networks.
\end{abstract}

\section{Introduction}

 Quantum correlations, particularly prior entanglement across quantum states, can be harnessed for transmitting more classical information through quantum communication links through teleportation and superdense coding schemes \cite{gisin2007quantum,mattle1996dense, pan2001entanglement, harrow2004superdense, vaidman1994teleportation}. The physical realization of quantum networked systems at atomic scale distances using such entangled quantum states is key towards realizing high-throughput quantum communications at such scales. The ideas from classical network coding, such as coding over butterfly networks, can be naturally extended to the quantum case, mindful of the quantum no-go theorems \cite{hayashi2007quantum,ahlswede2000network}. In the butterfly network, two units of information can
sent crossly and the channels can transmit only one bit, with the bottleneck being the central channel in the network. It was recently shown that perfect quantum state transmission is impossible in the butterfly network and the bottleneck, in the form of the central channel, cannot be resolved for a quantum network \cite{hayashi2007quantum}. This was subsequently extended to different kinds of networks with quantum network coding \cite{iwama2006quantum}. \textit{Leung et al} proposed network coding using shared entanglement between two parties \cite{leung2010quantum} through quantum teleportation and superdense coding. Hayashi demonstrated the impossibility of transmitting quantum states over the butterfly network between two senders without prior entanglement \cite{hayashi2007prior}. In this \textit{Letter}, we have formulated a non-trivial extension of Hayashi’s results to the case of higher-order butterfly networks using GHZ states. This extension is useful to realize polygon tesselated higher-order butterfly networks for recovery of network operability post detection of erased nodes. 

\section{Network Coding with GHZ States on higher-order Butterfly Networks}
Our protocol is based on a network with prior entanglement shared between three users. We fundamentally use teleportation for the proposed protocol. 
\begin{figure}[h]
\begin{center}
\includegraphics[width=8cm]{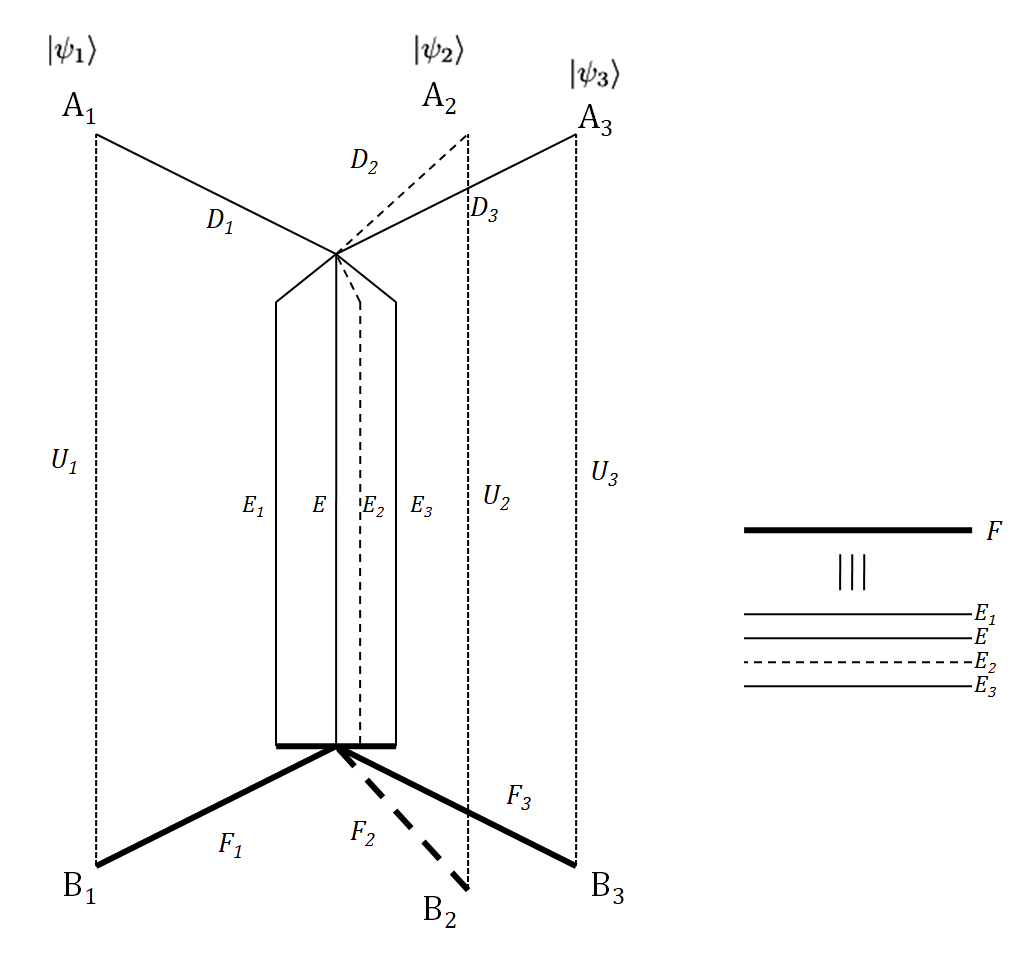}
\caption{Network for Quantum Butterfly Network comprising of three users with three GHZ states and an arbitrary quantum state for each transmission, alongwith four central channels: $E = (X_{1}^{(a)}\oplus X_{2}^{(a)} \oplus X_{3}^{(a)},X_{1}^{(b)}\oplus X_{2}^{(b)} \oplus X_{3}^{(b)})$, $E_1 = (X_{1}^{(a)}\oplus X_{2}^{(a)},X_{1}^{(b)}\oplus X_{2}^{(b)})$, $E_2 = (X_{2}^{(a)}\oplus X_{3}^{(a)},X_{2}^{(b)}\oplus X_{3}^{(b)})$ and $E_3 = (X_{1}^{(a)}\oplus X_{3}^{(a)},X_{1}^{(b)}\oplus X_{3}^{(b)})$.}
\label{Figure 1: Physical Realization}       
\end{center}
\end{figure}
We assume that the three senders $A_{1}$, $A_{2}$ and $A_{3}$ share three copies of maximally entangled (GHZ) states: $ \vert\phi_{i}\rangle = \frac{1}{\sqrt{2}}(\vert000\rangle + \vert111\rangle)_{(1,i),(2,i),(3,i)}$, where $i = 1, 2, 3$, denotes the $i^{\mathrm{th}}$ GHZ state, while the qubits with the same first index in the subscript belong to the same physical terminal. The senders prepare their states in: $\vert\psi_{j}\rangle = \alpha_{j}\vert0\rangle + \beta_{j}\vert1\rangle$, where $j = 1,2,3$ denotes the $j^{\mathrm{th}}$ user. In the first step, the sender $A_{i}$ performs a Bell state measurement $\{\phi_{+},\phi_{-},\psi_{+},\psi_{-}\}$ on the joint system $A_{i}\otimes A_{i,i}$. We can see the decomposition for $A_{i}\otimes A_{i,i}\otimes A_{(i+1) \mathrm{mod} 3,i}\otimes A_{(i+2) \mathrm{mod} 3,i}, i = 1,2,3$ in \textit{Table 1}. Here $A_{i}$ represents the arbitrary prepared state, while $A_{i,i}, A_{(i+1) \mathrm{mod} 3,i}$ and $A_{(i+2) \mathrm{mod} 3,i}$ are qubits from the three GHZ states at the physical terminal $i$. After the first Bell state measurement, we undertake an additional measurement on the single qubit component of the GHZ state at one of the other two nodes. We denote this measurement as $X_{i}^{(b)}$, with the first measurement being tagged as $X_{i}^{(a)}$. In this measurement, we measure single qubits in the $\vert\pm\rangle= \frac{1}{\sqrt{2}}(\vert0\rangle\pm\vert1\rangle$ basis.
\\
\\
The action of Bell state measurement leaves a state $U(X_{3}^{a},X_{2}^{b})^{-1}\vert\psi_{3}\rangle$ or $U(X_{2}^{a},X_{3}^{b})^{-1}\vert\psi_{2}\rangle$ depending on a sequence of measurements at the first qubit. Then the first terminal applies $U(X_{1}^{a},X_{1}^{b})$. The cumulative state can be represented by either of the following two cases
\begin{equation}
    U(X_{1}^{a},X_{1}^{b})U(X_{3}^{a},X_{2}^{b})^{-1}\vert\psi_{3}\rangle = c_{1132}U((X_{1}^{a},X_{1}^{b})\oplus (X_{3}^{a},X_{2}^{b}))^{-1}\vert\psi_{3}\rangle
\end{equation}
\begin{equation}
    U(X_{1}^{a},X_{1}^{b})U(X_{2}^{a},X_{3}^{b})^{-1}\vert\psi_{2}\rangle=c_{1123}U((X_{1}^{a},X_{1}^{b})\oplus (X_{2}^{a},X_{3}^{b}))^{-1}\vert\psi_{2}\rangle
\end{equation}
where $\vert c_{1123}\vert = \vert c_{1132}\vert = 1$, depending on whether terminal 2 or 3 measures $(a)$ or $(b)$. If we are to consider the clockwise cyclicity and the indices `wrapping around' (with $X_{4}=X_{1}$), we have the transformation $U((X_{i}^{a},X_{i}^{b})\oplus (X_{i+1}^{a},X_{i+2}^{b}))^{-1}$. 
To be able to recover the state at the receiving terminals, we formulate a combination of classical bits $X_{1}^{a}, X_{1}^{b}, X_{2}^{a}, X_{2}^{b}, X_{3}^{a}$ and $X_{3}^{b}$ to obtain an inverse unitary transformation for recovering the states. For the scheme with three users and three 3-qubit GHZ states, we use four central channels: $E = (\sum_{\oplus i=1}^{3}X_{i}^{(a)},\sum_{\oplus i=1}^{3}X_{i}^{(b)})$ and $E_{j} = (\sum_{\oplus i=1 \backslash j}^{3}X_{i}^{(a)},\sum_{\oplus i=1 \backslash j}^{3}X_{i}^{(b)})$. Classical information is transmitted using these four central channels to the three users, upon which the inverse transformation can be applied to the quantum state received at the terminal to get the cross-transmitted states. 
\begin{table}[h]
    \centering
    \begin{tabular}{|c|c|c|c|c|}
    \hline
       $X_{i}^{(a)}$ &  $X_{i}^{(b)}$ & $A_{i}\otimes A_{i,i}$ & $A_{(i+1) \mathrm{mod} 3,i}\otimes A_{(i+2) \mathrm{mod} 3,i}$ & $U$ \\ \hline
       (0,0)  & 0/1 & $\vert\psi_{+}\rangle$ & $\alpha_{1}\vert00\rangle+\beta_{1}\vert11\rangle$ & $I$/$\sigma_{z}$ \\ \hline
       (0,1)  & 0/1 & $\vert\psi_{-}\rangle$ & $\alpha_{1}\vert00\rangle-\beta_{1}\vert11\rangle$ & $\sigma_{z}$/$I$\\ \hline
       (1,0)  & 0/1 & $\vert\phi_{+}\rangle$ & $\alpha_{1}\vert11\rangle+\beta_{1}\vert00\rangle$ & $\sigma_{x}$/$\sigma_{x}\sigma_{z}$ \\ \hline
       (1,1)  & 0/1 & $\vert\phi_{-}\rangle$ & $\alpha_{1}\vert11\rangle-\beta_{1}\vert00\rangle$ & $\sigma_{x}\sigma_{z}$/$\sigma_{x}$\\ \hline
    \end{tabular}
    \caption{Decomposition for $ A_{i}\otimes A_{i,i}\otimes A_{(i+1)  \mathrm{mod} 3,1}\otimes A_{(i+2) \mathrm{mod} 3,1} \forall i =1,2,3$, and truth values as well as associated unitary transformations for higher-order butterfly network over three nodes}
    \label{tab:my_label}
\end{table}
This proposal can be generalised to the case of $n$-qubits by considering an $n$-qubit GHZ state: $\vert\xi\rangle = \frac{1}{\sqrt{2}}(\vert0^{\otimes n}+\vert1^{\otimes n})$. In this case, there will be $n$+1 channels with $E = (\sum_{\oplus i=1}^{n}X_{i}^{(a)},\sum_{\oplus i=1}^{n}X_{i}^{(b)})$ and $E_{j} = (\sum_{\oplus i=1 \backslash j}^{n}X_{i}^{(a)},\sum_{\oplus i=1 \backslash j}^{n}X_{i}^{(b)})$. As in the case of the two parties, perfect transmission of quantum states is impossible without prior entanglement. Following the analysis of \textit{Hayashi} \cite{hayashi2007prior} for the case of EPR state, we find the bound on fidelity for a generalised case to be 
\begin{equation}
    \sum_{i=1}^{d}f_{i}\leq \frac{2.8512 d}{d+1}
\end{equation}
where $f_{i}$ is the average entanglement fidelity of the $i^{\mathrm{th}}$ channel and $d$ is the number of nodes in the quantum network. The operation of this higher-order butterfly scheme cannot be undertaken without prior entanglement. 
\section{Scheme for Quantum Network Recovery from Multinode Failure using Tesselated GHZ-Butterfly Subnetworks}. 
The $n$-qubit GHZ states can be used for network recovery after multinode failure for an arbitrary graph state using tesselation of such GHZ-states in partitioned blocks of the graph, as shown in \textit{Figure 2 (a)}. The scheme for undertaking network recovery using the higher-order GHZ state based network coding, formulated in the previous section, comprises of two distinct steps: Checking for operability of nodes and recovery of network operability. We prepare the graph state that constitutes the quantum network by initialising the states at the first terminal/node $\vert\phi\rangle$ and the other nodes with the state $\vert+\rangle=\frac{1}{\sqrt{2}}(\vert0\rangle+\vert1\rangle)$. We then operate with the \textit{CPHASE} operation: $CZ_{ij}=\vert0\rangle\langle0\vert_{i}I_{j}+\vert1\rangle\langle1\vert_{i}Z_{j}$, where $Z$ is the Pauli-z matrix, between adjacent nodes. This gives us the entangled graph state which is the primary resource in the quantum network being studied. 
\\
\\
For checking for operability of nodes, we use a hybrid classical-quantum approach. Classically, communication of node failure is done using `pings' that are used in the ICMP echo protocol. However, this is not enough to ascertain quantum accessibility and entanglement at the node. Additional to the classical connection, we need an additional layer to enable possibility to assess whether entanglement at the node is accessible and operative. This is done using quantum non-destructive measurements on the nodes of the graph, and any instance of node failure is relayed to the rest of the nodes in the affected subgraph using  classical infrastructure. 
For optical systems, we can utilise non-destructive detection using an optical resonator containing a single atom prepared in
a superposition of two states \cite{reiserer2014quantum}.
\\
\begin{figure*}[htb!]
\begin{center}
\includegraphics[width=15cm]{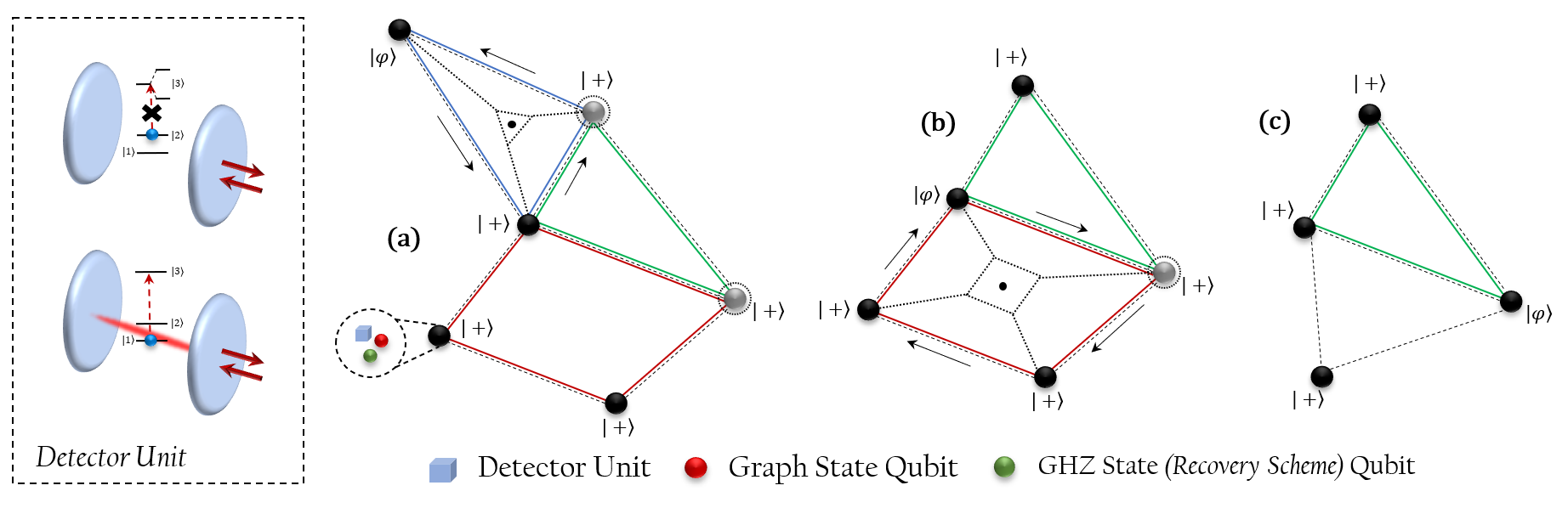}
\caption{Illustrative example for scheme for recovery of network operability from multinode failure: (a) The nodes begin with $\vert\phi\rangle$ at one node and $\vert+\rangle$ at the others, and we apply C-Phase operations on all the adjacent nodes. We then consider the possibility of two nodes failing (marked in grey). The coloured lines represent the GHZ states shared between adjacent nodes. (a)-(c) Mechanism for recovery upon multinode recovery of the network, using GHZ-based network coding in segments of the composite network. Inset is the detector unit comprising of an optical resonator with a three-level atom in a single-sided cavity.}
\label{Figure 1: Physical Realization}       
\end{center}
\end{figure*}
\\
At each node, we position an optical resonator system with a three-level atom in a single-sided cavity, where one of the two mirrors is perfectly reflecting and there is a small transmission coefficient of the other mirror that allows for in- and outcoupling of light, which is an optical photon. Let the three states of the atom be tagged $\vert1_{a}\rangle$, $\vert2_{a}\rangle$ and $\vert3_{a}\rangle$. The cavity is designed such that it is overcoupled and resonant with the transition between the atomic states $\vert2_{a}\rangle$ and $\vert3_{a}\rangle$. If we prepare the atom initially in the state $\frac{\vert1_{a}\rangle+\vert2_{a}\rangle}{\sqrt{2}}$, an impinging photon makes it transform to $\frac{\vert2_{a}\rangle-\vert1_{a}\rangle}{\sqrt{2}}$ while it remains unchanged in the absence of an impinging photon. This is due to there being no interaction between the atom and photon when the atom is in the state $\vert1_{a}\rangle$ since any transition is far detuned, while when the atom is in the state $\vert2_{a}\rangle$, the strong photon-atom coupling can lead to a normal-mode splitting and the photon undergoes reflection without entering the cavity. We can measure this phase flip using a $\frac{\pi}{2}$ rotation map to map $\frac{\vert1_{a}\rangle+\vert2_{a}\rangle}{\sqrt{2}} \rightarrow \vert1_{a}\rangle$ and $\frac{\vert1_{a}\rangle-\vert2_{a}\rangle}{\sqrt{2}} \rightarrow \vert2_{a}\rangle$, and then cavity-enhanced fluorescence state detection can be used to distinguish between the states $\vert2_{a}\rangle$ and $\vert1_{a}\rangle$ \cite{bochmann2010lossless}. 
\\
\\
Upon the non-destructive detection of a lost photon, the classical infrastructure used for classical communication in our model for higher-order butterfly networks is used to alert the other nodes in the subnetworks of which the failed node(s) is(are) a part of. This is undertaken using a parity code check. This comprises of $E = (\sum_{\oplus i=1}^{n}X_{i}^{(a)},\sum_{\oplus i=1}^{n}X_{i}^{(b)})$ and $E_{j} = (\sum_{\oplus i=1 \backslash j}^{n}X_{i}^{(a)},\sum_{\oplus i=1 \backslash j}^{n}X_{i}^{(b)})$ for $n$ nodes of a subnetwork. However, unlike in the case of the higher-order butterfly network, we do not send any single- or two-qubit measurements over these channels but rather if the check on each node yields a success (`1') or failure (`0'). In this manner, each node in the subnetwork can exactly know which node(s) is(are) non-operative. Upon ascertaining the same, we operate with a $\sigma_{z}$ operate to remove the particular failed node(s) from the graph state. 
\\
\\
For recovery of network operability, we can use the network coding formalism for higher-order butterfly networks developed and discussed in the previous section. In this framework, the key point is to tesselate the complex multiqubit network with subnetworks comprising of $n_{i}$ qubit GHZ state for the $i^{\mathrm{th}}$ subnetwork comprising of $n_{i}$ nodes, along with the requisite classical infrastructure, as highlighted in the model formulated for higher-order butterfly networks. As soon as we ascertain which nodes are non-operative, we can use the GHZ-state in the subnetwork to replace the failed node, as shown in \textit{Figure 2 (a)-(c)}. Failure of all nodes that a subnetwork shares with the rest of the network, along with both additional qubits on either side of this set precipitates a point of criticality wherein the network operability cannot be recovered using the formulated scheme. The recovery of rate-optimal coded graph state networks can also be realised using \textit{modified graph-state codes} \cite{priya}. 
\section{Conclusion} In this \textit{Letter}, we have formulated and constructed a higher-order butterfly network using a multiqubit GHZ states. We studied the fidelity bounds of transmission in such GHZ-based higher-order butterfly networks that scale as $\frac{n}{n+1}$ for the number of terminals, which we use to construct a scheme for quantum network recovery from multinode failure. This is particularly useful for making quantum networks resilient against failures at multiple nodes. 
\section{Acknowledgement}
We would like to acknowledge Prof. Brian Josephson (University of Cambridge) for his comments on the \textit{Letter}.
\bibliographystyle{alpha}
\bibliography{main}

\end{document}